\documentstyle[preprint,aps,floats,epsfig]{revtex}

\begin{document}

\draft

\preprint{\rightline{ANL-HEP-PR-96-56 \hspace{.5in} FSU-SCRI-96-139}}

\title{THERMODYNAMICS OF LATTICE QCD WITH TWO LIGHT QUARK FLAVOURS ON A
       $16^3 \times 8$ LATTICE II}

\author{S.~Gottlieb}
\address{Department of Physics, Indiana University, Bloomington, IN 47405, USA}
\author{U.~M.~Heller and A.~D.~Kennedy}
\address{SCRI, The Florida State University, Tallahassee, FL 32306-4052, USA}
\author{S.~Kim} 
\address{Center for Theoretical Physics, Seoul National University, Seoul, 
Korea}
\author{J.~B.~Kogut}
\address{Department of Physics, University of Illinois, 1110 West Green Street,
Urbana, IL 61801, USA}
\author{C.~Liu}
\address{Morgan Stanley \& Co. Inc., 1585 Broadway, New York, NY 10036, USA}
\author{R.~L.~Renken}
\address{Department of Physics, University of Central Florida, Orlando, FL
32816, USA}
\author{D.~K.~Sinclair}
\address{HEP Division, Argonne National Laboratory, 9700 South Cass Avenue,
Argonne, IL 60439, USA}
\author{R.~L.~Sugar}
\address{Department of Physics, University of California, Santa Barbara, CA
93106, USA}
\author{D.~Toussaint}
\address{Department of Physics, University of Arizona, Tucson, AZ 85721, USA}
\author{K.~C.~Wang}
\address{China Graduate School of Theology, 5 Devon Rd, Kowloon Tong, Kowloon,
Hong Kong}

\maketitle

\begin{abstract}
We have extended our earlier simulations of the high temperature behaviour of
lattice QCD with two light flavours of staggered quarks on a $16^3 \times 8$
lattice to lower quark mass ($m_q=0.00625$). The transition from hadronic matter
to a quark-gluon plasma is observed at $6/g^2=5.49(2)$ corresponding to a 
temperature of $T_c \approx 140$MeV. We present measurements of observables
which probe the nature of the quark-gluon plasma and serve to distinguish it 
from hadronic matter. Although the transition is quite abrupt, we have seen no
indications that it is first order.
\end{abstract}


\setcounter{page}{1}
\pagestyle{plain}
\parskip 5pt
\parindent 0.5in

\section{INTRODUCTION}

The understanding of the thermodynamics of hadronic matter is a major goal of
lattice QCD. In particular, one would like to find and understand the nature
of the transition from hadronic matter to a quark-gluon plasma and to determine
the properties of this plasma phase which serve to distinguish it from the
hadronic phase. Such high temperature matter must surely have existed in the
early universe, and its properties would have helped determine the evolution
of the universe. In addition, it is believed that heavy-ion collisions such as
will be observed at RHIC will produce hot nuclear/hadronic matter and probably
quark-gluon plasma. Finally, such studies can help us understand the dynamics
of QCD including confinement and chiral symmetry breaking. 

We have carried out simulations of QCD with 2 flavours of light staggered
quarks on a $16^3 \times 8$ lattice. In the first paper in this series we
described simulations with quark mass $m_q=0.0125$, in lattice units
\cite{part1}. In the current paper we describe simulations with quark mass
$m_q=0.00625$ and, where possible, attempt to extrapolate to the chiral limit.
Prior results of this study were reported at Lattice conferences
\cite{amsterdam,bielefeld}, and there has also been a  report from the Columbia
group on their work with a lighter quark mass, $0.004$ \cite{mawhinney}.

The simulations were performed on the Connection Machine (CM-2) at the
Pittsburgh Supercomputing Center. We used the hybrid molecular dynamics
algorithm with noisy fermions for our simulations enabling us to tune the
number of quark flavours to 2. As at the higher mass, we found no evidence for
a first-order transition, in agreement with work on lattices with $N_t=4$
\cite{4}, $6$ \cite{6}, and $12$ \cite{12}. This is what was predicted on the
basis of chiral spin models, which predict a second-order transition at $m_q=0$
and a rapid crossover, with no transition at small $m_q$ \cite{pisarski}.

The position of the transition was determined by measuring the Wilson/Polyakov
line, which is associated with confinement, and the chiral condensate,
$\langle\bar{\psi}\psi\rangle$, which measures chiral symmetry breaking.
In addition, we have calculated the entropy densities of both quarks and gluons,
and measured the topological susceptibility. From the correlation functions
of spatial and temporal Wilson lines, we have estimated the string tensions and
Debye screening lengths, and addressed the question of $\psi/J$ production.
By measuring the baryon number susceptibility, we have determined how the
system responds to a finite chemical potential for baryon number. Additionally,
we have measured the hadronic screening lengths for light hadronic excitations.
These should give us an indication as to whether the plasma has any hadron-like
excitations or whether its excitations are simply quarks and gluons. Finally,
we have added valence ``strange'' quarks, whose masses are tuned to give the
correct kaon masses when combined with $u$ or $d$ quarks of mass $m_q=0.00625$.
By measuring the entropy of such strange quarks, we can address the question
as to whether the formation of the quark-gluon plasma is characterized by
increased kaon production. We have also measured the kaon screening lengths.

Section 2 describes the simulations and the extractions of most of the order
parameters. In section 3 we discuss Debye and hadronic screening lengths.
Finally, in section 4 we present further discussions and conclusions. 

\section{SIMULATIONS}

These simulations were performed on a $16^3 \times 8$ lattice with 2 flavours
of light quarks with mass $m_q=0.00625$, which will be compared, when
appropriate, with earlier simulations with $m_q=0.0125$ \cite{part1}. The
updating was performed with time increment $dt=0.005$, except at $6/g^2=6.0$
and the first half of the simulations at $6/g^2=5.55$ where it was chosen to be
$dt=0.01$. We have simulated for 1000 time units at $6/g^2=5.45$, 1500 time
units at $6/g^2=5.475$, 1145 time units from a cold start and 1000 time units
from a hot start at $6/g^2=5.5$, and 1000 time units at each of $6/g^2=5.525$,
$5.55$ and $6.0$. Configurations were saved every 20 time units at $6/g^2=6.0$,
every 10 time units at $6/g^2=5.55$, and every 5 time units for each of the
other couplings. These configurations were used for later analyses. 

\begin{figure}[htb]
\epsfxsize=4in
\centerline{\epsffile{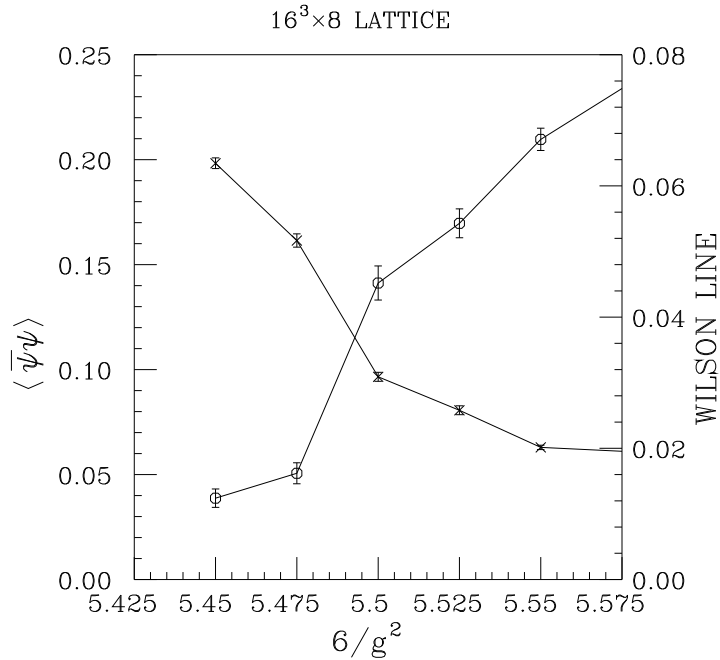}}
\caption{The Wilson/Polyakov line (circles) and $\langle\bar{\psi}\psi\rangle$
(crosses) as functions of $6/g^2$.\label{fig:wil-psi}}
\end{figure}

\begin{figure}[htb]
\centerline{\hbox{\psfig{figure=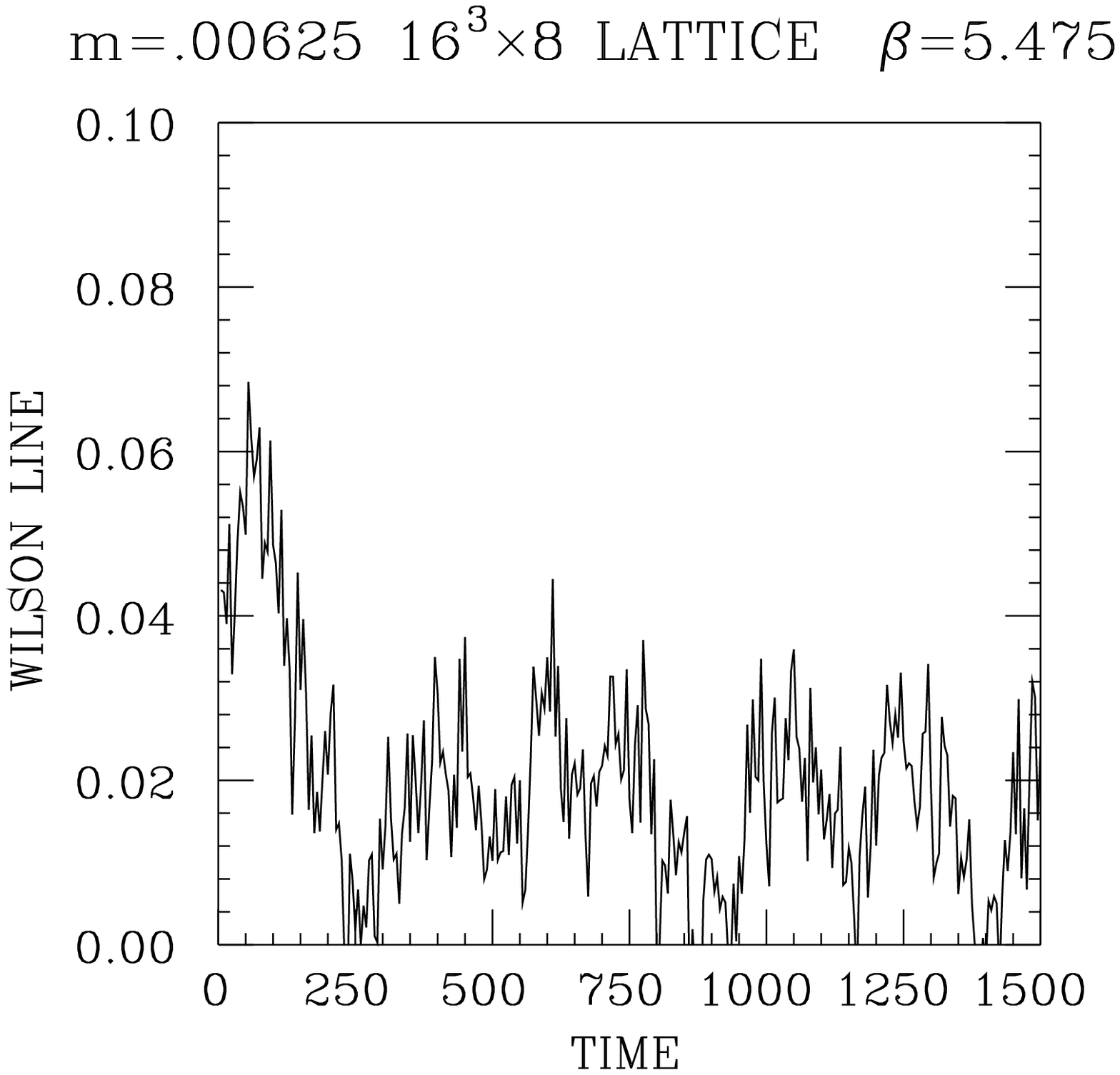,height=9cm,width=9cm}
                  \psfig{figure=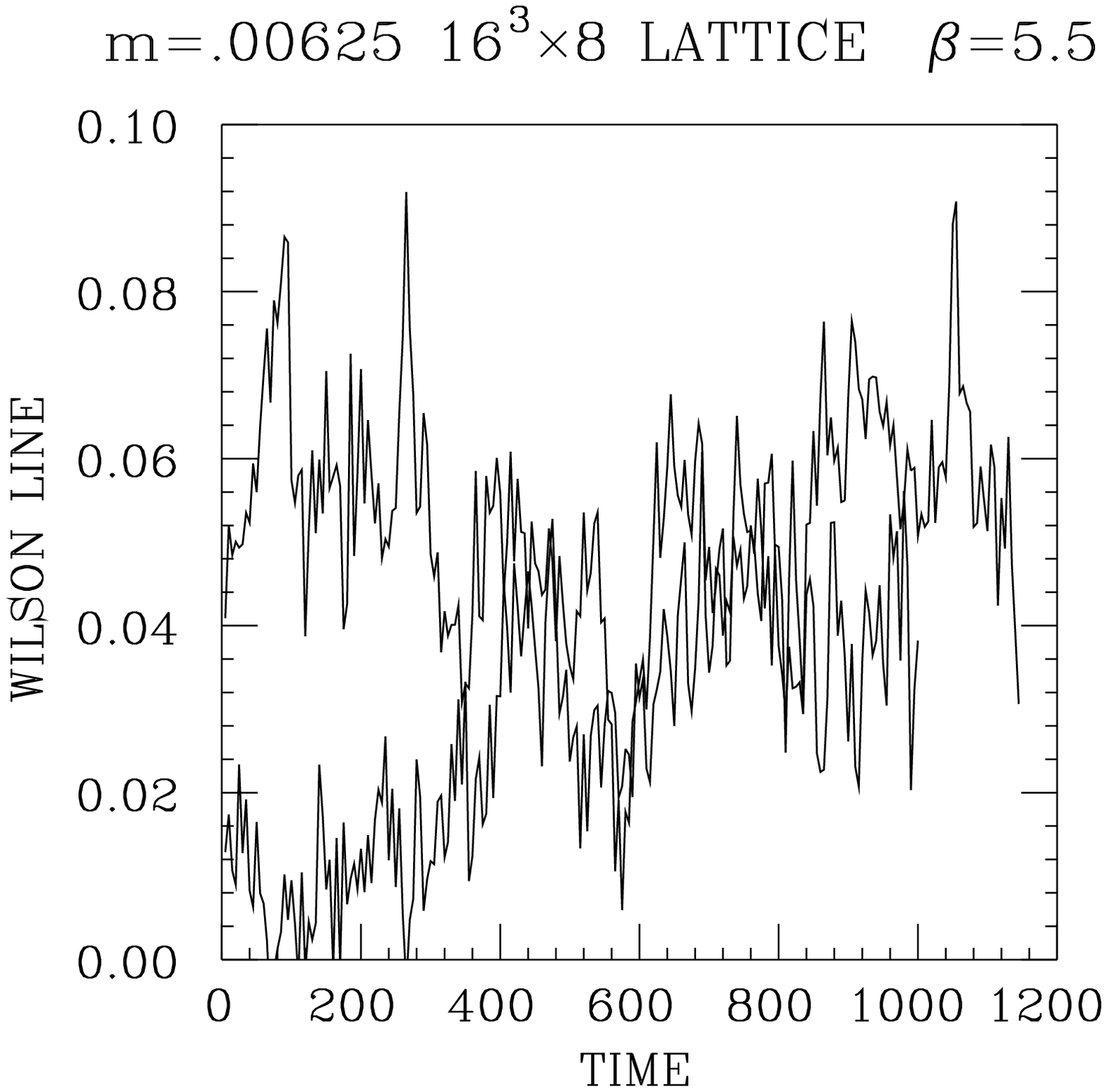,height=9cm,width=9cm}}}
\centerline{(a) \hspace{2.5in} (b)}
\vspace{0.1in}
\caption{Time evolution of the Wilson/Polyakov line (a) at $6/g^2=5.475$
         and (b) at $6/g^2=5.5$.\label{fig:wilson}}
\end{figure}

The thermal(temporal) Wilson/Polyakov line, and the chiral condensate 
$\langle\bar{\psi}\psi\rangle$ were measured and are plotted in
Fig.~\ref{fig:wil-psi}. They are tabulated, along with values of the plaquette,
in Table~\ref{tab:wil-psi}. In each case the first 200 time units are discarded
for equilibration. The rapid crossover of these two quantities between
$6/g^2=5.475$ and $6/g^2=5.5$ suggests that the transition occurs in this
range. This is further born out by looking at the time histories of the
Wilson/Polyakov line at $6/g^2=5.475$ and $5.5$. These are shown in
Fig.~\ref{fig:wilson}. At $6/g^2=5.5$, starts from both the hot and cold sides
of the transition evolve to the high-temperature phase. At $6/g^2=5.475$,
starting from the high temperature phase, the system rapidly settles into the 
low temperature phase and stays there. 
\begin{table}[htb]
\begin{tabular}{|l|d|d|d|}
$6/g^2$      &  Wilson Line  &  $\langle\bar{\psi}\psi\rangle$ & plaquette \\
\hline
5.45         &  0.0124(14)   &   0.1983(25)   &   0.45780(11)  \\     
5.475        &  0.0162(16)   &   0.1615(32)   &   0.45328(12)  \\
5.5          &  0.0452(26)   &   0.0966(21)   &   0.44850(9)   \\
5.525        &  0.0543(22)   &   0.0806(21)   &   0.44493(9)   \\
5.55         &  0.0671(17)   &   0.0630(10)   &   0.44099(7)   \\
6.0          &  0.2067(24)   &   0.02925(3)   &   0.39243(3)   \\
\end{tabular}
\caption{The Wilson/Polyakov line and $\langle\bar{\psi}\psi\rangle$ as
functions of $6/g^2$ for $m_q=0.00625$}
\label{tab:wil-psi}
\end{table}

\begin{figure}[htb]
\centerline{\hbox{\psfig{figure=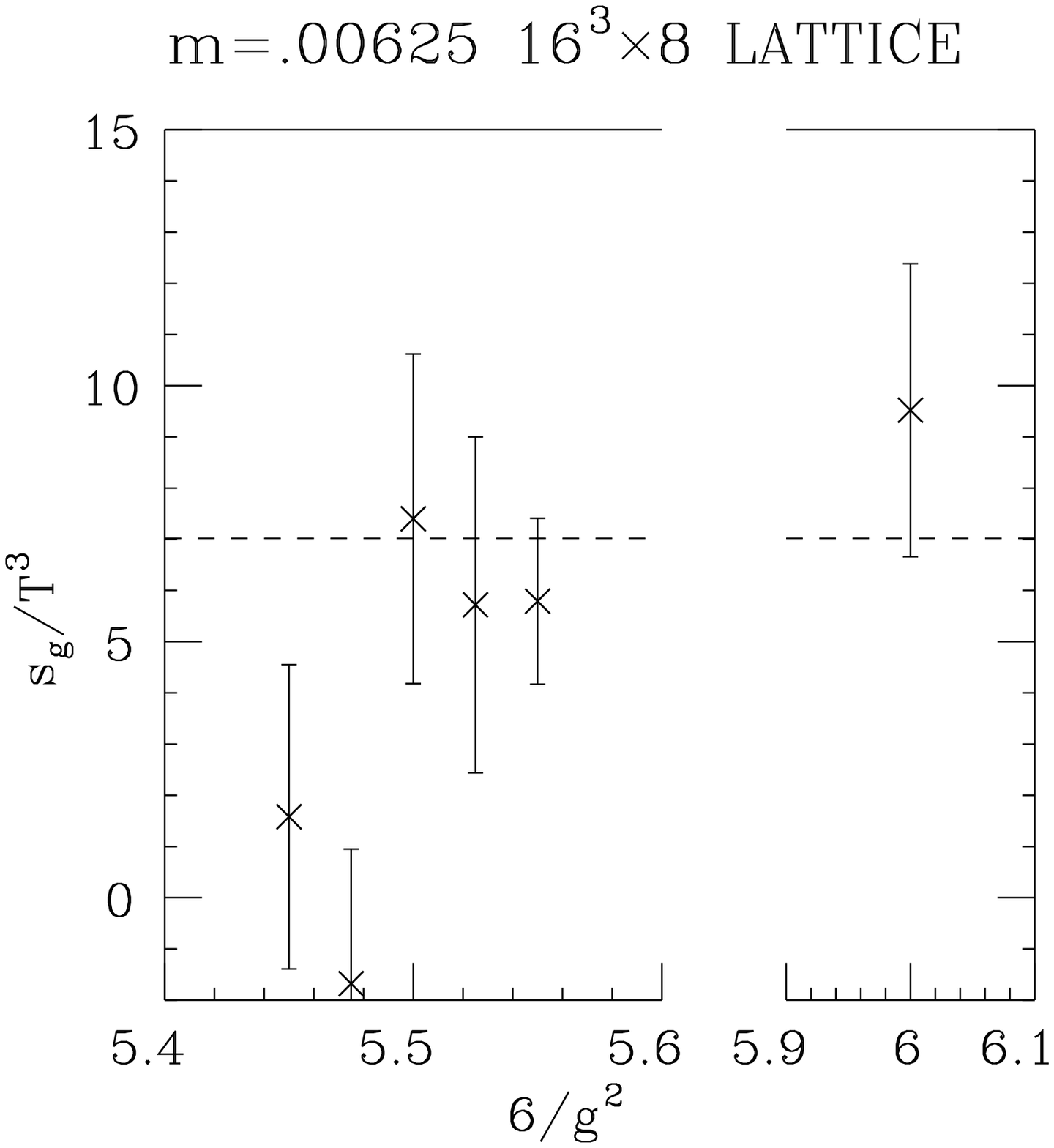,height=8cm,width=8cm}
                  \psfig{figure=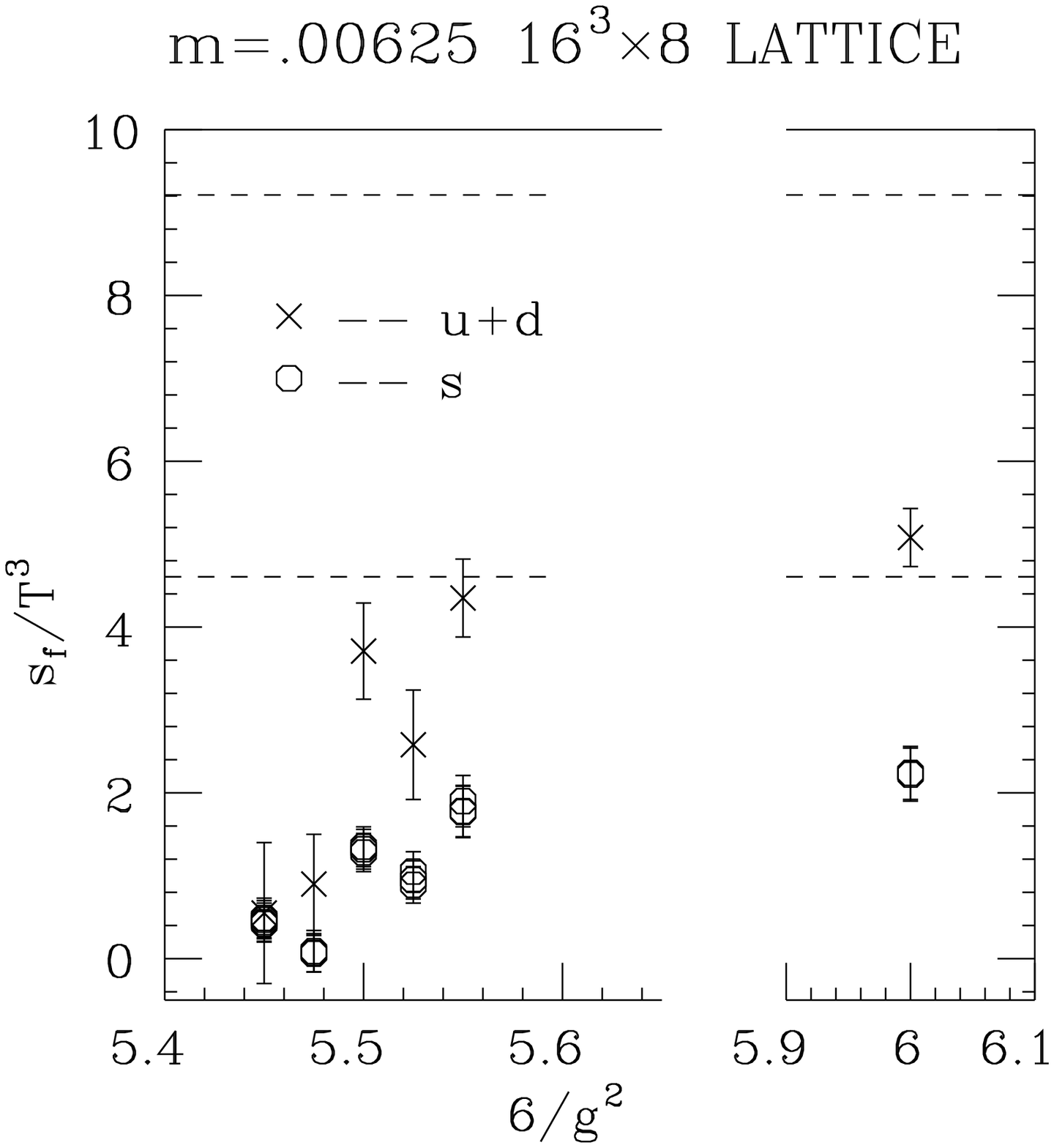,height=8cm,width=8cm}}}
\centerline{(a) \hspace{2.5in} (b)}
\vspace{0.1in}
\caption{Partial entropy densities of (a) the gluons and (b) the quarks as 
function of $6/g^2$. The dashed lines are the Stefan-Boltzmann values, in
(b) the lower dashed line is for the $s$ quarks and the upper is for the 
$u + d$ quarks. \label{fig:entropies}} 
\end{figure}

We have measured the partial entropy densities for the gluons and the light
dynamical ($u$ and $d$) quarks. In addition, we have calculated the partial
entropy densities for a heavier ``strange''($s$) valence quark. At tree order
the partial entropies for the quarks and gluons are given by 
\begin{equation}
     s^0_g T = \frac{4}{g^2}(P_{st} - P_{ss})
\end{equation}
and
\begin{equation}
     s^0_f T = \frac{n_f}{3}\left[\langle\bar{\psi}D\!\!\!\!/_0\psi\rangle 
              -\frac{3}{4}+\frac{1}{4}m_q\langle\bar{\psi}\psi\rangle\right]
\end{equation}
respectively, where $P_{st}$ and $P_{ss}$ are the space-time and space-space
plaquettes normalized as in \cite{part1}. Since at this order, the energy
density $\epsilon$ and pressure $p$ obey the perfect gas relation 
$p=\frac{1}{3}\epsilon$, which is clearly false for a first order transition,
we include the 1-loop corrections. Through 1-loop order \cite{karsch},
\begin{equation}
     s_g T = \left(1-\frac{1.279}{6/g^2}\right) s^0_g T
\end{equation}
\begin{equation}
     s_f T = \left(1-\frac{1.022}{6/g^2}\right) s^0_g T
\end{equation}
The tree level quantities are tabulated in Table~\ref{tab:entropies}. The
multiple values for the strange quark at each $\beta$ correspond to different
choices for the strange quark mass, which span the range of estimates for the
masses required to give the correct kaon mass when $m_{u,d}=0.00625$. The
values chosen were 0.035, 0.040, 0.045 and 0.050 at $\beta=5.45$, 0.030, 0.035
and 0.040 at $\beta=5.475$, 0.025, 0.030 and 0.035 at $\beta=5.5$, 0.020, 0.025
and 0.030 at $\beta=5.525$, 0.0175, 0,0225, 0.0275 at $\beta=5.55$ and 0.010
and 0.015 at $\beta=6.0$.  
\begin{table}[htb]
\begin{tabular}{|l|d|d|d|d|d|d|}
$\beta$  & $s^0_g/T^3$ & $s^0_{u+d}/T^3$ & \multicolumn{4}{c|}{$s^0_s/T^3$} \\
\hline 
5.45  &  1.9(3.6) &  1.2(1.8) & 1.0(5) & 1.0(5) & 0.9(5) & 0.9 (5) \\
5.475 & -2.1(3.2) &  1.9(1.2) & 0.2(5) & 0.2(5) & 0.1(5) &         \\
5.5   &  9.0(3.9) &  7.7(1.2) & 2.8(5) & 2.7(5) & 2.6(5) &         \\
5.525 &  7.0(4.0) &  5.0(1.4) & 2.1(5) & 2.0(5) & 1.8(5) &         \\
5.55  &  7.0(3.5) &  8.8(0.9) & 3.9(6) & 3.7(6) & 3.7(6) &         \\
6.0   & 11.3(3.4) & 10.2(0.7) & 4.5(7) & 4.5(7) &        &
\end{tabular}
\caption{Partial entropy densities for gluons, $u$ plus $d$ quarks, and $s$ 
quarks at tree order.}
\label{tab:entropies}
\end{table}
The 1-loop corrected partial entropy densities, each divided by the ratio of
the corresponding free field quantity on a $16^3 \times 8$ lattice to its
continuum value, in an attempt to remove some of the finite size and finite
lattice spacing effects
\cite{engels}, are plotted in Fig.~\ref{fig:entropies}. The above ratios are
0.99136 for gluons and 1.59828 for $u$ and $d$ quarks. The gluon partial
entropy density divided by $T^4$ is small for small values of $6/g^2$, and
increases abruptly to close to the Stefan-Boltzmann value ($32\pi^2/45$)
between $6/g^2=5.475$ and $6/g^2=5.5$, and remains there as $6/g^2$ is
increased. The partial entropy for the $u + d$ quarks over $T^3$ also increases
rapidly between $6/g^2=5.475$ and $6/g^2=5.5$, and slowly increases for higher
values. Even by $6/g^2=6.0$, it is still little more than half its
Stefan-Boltzmann value ($7\pi^2 n_f/15$). This slow approach to its asymptotic
value could well be due to poor convergence of the perturbative loop expansion.
However, in contrast to the case of Wilson fermions, tadpole improvement is not
expected to be (and in fact is not) of any help \cite{mackenzie}. The entropy
density of the strange quark behaves in a fashion qualitatively similar to that
of the $u+d$ quarks. Just above the transition (at $6/g^2=5.5$) it is about
36\% of that of the $u+d$ quarks, while by $6/g^2=6.0$ it has risen to about
44\% on its way to the high temperature limit of 50\%. 

\begin{figure}[htb]
\epsfxsize=4in
\centerline{\epsffile{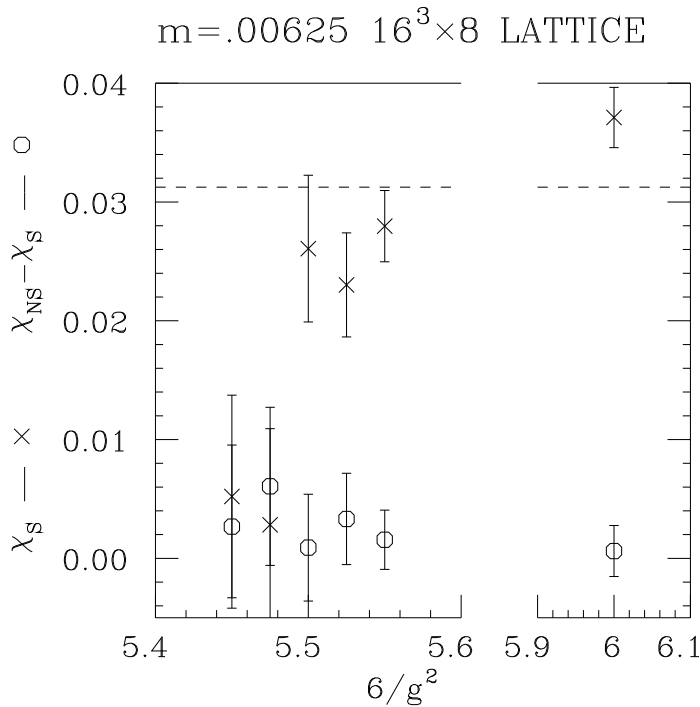}}
\caption{Quark number susceptibility $\chi_S$ and $\chi_{NS}-\chi_S$ as 
functions of $6/g^2$.\label{fig:baryon}}
\end{figure}

For a complete understanding of the thermodynamics of lattice QCD one really
needs to work at finite baryon- and hence quark-number density. Although that
is beyond the scope of these simulations, we have measured the singlet(S) and
non-singlet(NS) quark-number susceptibilities \cite{bns},
\begin{equation}
\chi_S = \left(\frac{\partial}{\partial\mu_u}+\frac{\partial}{\partial\mu_d} 
         \right)(\langle n_u \rangle + \langle n_d \rangle)
\end{equation}
and
\begin{equation}
\chi_{NS} = \left(\frac{\partial}{\partial\mu_u}-\frac{\partial}{\partial\mu_d} 
            \right)(\langle n_u \rangle - \langle n_d \rangle)
\end{equation}
where $\langle n_u \rangle$ and $\langle n_d \rangle$ are the expectation
values of the number densities for $u$ and $d$ quarks and $\mu_u$ and $\mu_d$
are chemical potentials for $u$ and $d$ quarks. We have evaluated $\chi_S$ and
$\chi_{NS}$ on the lattice evaluated at zero chemical potentials using the
method of \cite{bns}   with 50 noise vectors (except at $6/g^2=5.55$ where we
used 20, 50, and 100 noise vectors -- we report results only for 100 noise
vectors -- which indicated that 50 noise vectors was adequate), and plotted our
results in Fig.~\ref{fig:baryon}. The difference between $\chi_{NS}$ and
$\chi_S$ remains small throughout the range. $\chi_S$ is small below the
transition, rising rapidly to close to the continuum limit -- $N_f T^2 =
0.03125$ -- as $6/g^2$ is increased through the transition, exceeding the
continuum value for large $6/g^2$, a possible finite size effect. The singlet
susceptibility measures the ease with which a quark excess can be created by a
finite chemical potential. Below the transition, confinement prevents the
production of quarks, except as baryons, which requires extra energy: thus
$\chi_S$ is small. Above the transition unconfined quarks can be created, which
requires little energy, so $\chi_S$ is large. $\chi_{NS}$ measures the ease
with which an excess of $u$ quarks and anti-$d$ quarks can be created. Below
the transition, the easiest way to do this is to create an excess of $\pi^+$'s.
However, the mass of the pion (larger than normal since $m_q$ is larger than
its physical value) tends to suppress such production, and $\chi_{NS}$ is
small. Above the transition $u$ and $\bar{d}$'s can be created with little
energy and $\chi_{NS}$ is large. 

Quark number susceptibility is just one of the quantities which is of 
importance if our eventual goal is to understand the equation of state near
the transition. Other such quantities are the fluctuation quantities such as
\cite{kl,bktg}
\begin{equation}
C_V=\frac{1}{VT^2}\frac{\partial^2}{\partial(1/T^2)}{\rm ln}Z
\end{equation}
\begin{equation}
\chi_m=\frac{T}{V}\frac{\partial^2}{\partial m_q^2} {\rm ln}Z
\end{equation}
\begin{equation}
\chi_t=\frac{T}{V}{\partial^2 \over \partial m_q \partial(1/T)} {\rm ln}Z
\end{equation}
Since such quantities have peaks which span only a small range of $6/g^2$, to
calculate them directly requires knowing their values at $6/g^2$'s which are
closer together than typically simulated. This deficiency can often be overcome
by interpolating using the Ferrenberg-Swendsen \cite{fs}
method. We found, unfortunately,
that the spacing of our simulated $6/g^2$'s -- $\Delta 6/g^2=0.025$ in the 
transition region -- was too large to permit such an interpolation. We estimate
that one would need a $\Delta 6/g^2 \approx 0.01$ (or less) to enable such an 
interpolation on this size lattice (not unreasonable for future simulations).
The contributions to these quantities at only the simulated values of $6/g^2$ 
show little variation, and no peaks -- the transition region is too narrow
and none of our points are close enough to feel much effect. In the case of
$\chi_m$, since only one noise vector was used for each estimate of 
$\langle\bar{\psi}\psi\rangle$, it is probable that the fluctuations from use
of this stochastic estimator completely overwhelmed those associated with the
gauge configurations \cite{lagae_pc}.

From our measurements of the Wilson/Polyakov Line, 
$\langle\bar{\psi}\psi\rangle$, and entropy density, we conclude that the
transition from hadronic matter to a quark-gluon plasma occurs at 
$6/g^2=5.49(2)$ for $m_q=0.00625$ as compared to $6/g^2=5.54(2)$ for 
$m_q=0.0125$. Linear extrapolation leads us to predict that the transition
would occur at $6/g^2=5.44(3)$ in the chiral ($m_q=0$) limit. This $m_q$
dependence of the position of the phase transition makes it difficult to
extrapolate observables to $m_q=0$, since to extrapolate to $m_q=0$ at constant
$6/g^2$, requires $6/g^2$ to lie on the same side of the transition at $m_q=0$
as at $m_q=0.0125$. Of course, one might argue that it is more physical to
continue, not at constant $6/g^2$ but rather at constant physics. Since the
change in $6/g^2$ for the transition $\Delta(6/g^2) \approx 0.05$ between
$m_q=0.00625$ and $m_q=0.0125$ and between $m_q=0$ and $m_q=0.00625$, we have
attempted linear extrapolations in mass with $\Delta(6/g^2)=0.05$ between
$m_q=0.00625$ and $m_q=0.0125$ as well as for the traditional
$\Delta(6/g^2)=0$. In addition, we have included extrapolations for
$\Delta(6/g^2)=0.025$. Such extrapolations are shown for the chiral condensate
$\langle\bar{\psi}\psi\rangle$ in Table~\ref{tab:chiral}. Except at
$6/g^2=5.45$ we have restricted ourselves to those for which the $m_q=0.0125$,
the $m_q=0.00625$ and $m_q=0$ lie on the same side of the transition. At
$6/g^2=5.45$ we have included cases where the $m_q=0.0125$ and $m_q=0.00625$
points lie in the hadronic phase while the $m_q=0$ point (probably) lies in the
plasma phase. Although not representing the correct chiral limit, these, we
believe, are the correct values to use to predict the topological charge. 
\begin{table}[htb]
\begin{tabular}{|l|d|d|d|}
       & \multicolumn{3}{c|}{$\langle\bar{\psi}\psi\rangle$}               \\
$6/g^2$ & \multicolumn{1}{c|}{$\Delta(6/g^2)=0.000$} 
        & \multicolumn{1}{c|}{$\Delta(6/g^2)=0.025$}
        & \multicolumn{1}{c|}{$\Delta(6/g^2)=0.050$} \\
\hline
5.400  &                    &                    & 0.2161(56)        \\
5.425  &                    &                    & 0.1660(67)        \\
5.450  &  0.1420(58)$^\ast$ &  0.1425(68)$^\ast$ & 0.0785(44)        \\ 
5.475  &                    &                    & 0.0584(43)        \\
5.500  & [0.0126(49)]       &  0.0467(23)        & 0.0285(24)        \\
5.525  & [0.0043(47)]       &  0.0230(23)        &                   \\
5.550  &  0.0113(23)        &                    &                   \\
6.000  &  0.00011(9)        &                    &                   \\
\end{tabular}
\caption{$\langle\bar{\psi}\psi\rangle$ extrapolated to $m_q=0$. Those values
         marked with an asterisk are not to be interpreted as true $m_q=0$ 
         values, but are included because they are related to topological 
         charge. Naive extrapolations which cross the transition are included  
         in square brackets for comparison.}
\label{tab:chiral}
\end{table}
It is amusing to note that, above the transition (at $m_q=0$), where
$\langle\bar{\psi}\psi\rangle$ should vanish at $m_q=0$, the extrapolations at
constant $6/g^2$, even where they cross the transition, do better than those
where $6/g^2$ is changed to avoid the transition. 

We have calculated the topological charge $Q$ on each configuration using the
cooling method \cite{teper} described in more detail in our earlier paper 
\cite{part1}. From these $Q$'s
we have calculated the topological susceptibilities $\chi$ as
\begin{equation}
           \chi = \frac{1}{V}\langle Q^2 \rangle
\end{equation}
The values we obtain by applying this method naively at $6/g^2=5.45$ are
actually slightly larger at this quark mass ($m_q=0.00625$) than they were at
$m_q=0.0125$ ($28(3) \times 10^{-5}$ compared with $24(5) \times 10^{-5}$) whereas $\chi$ should be 
proportional to $m_q$ and thus smaller at the lower mass. For this reason we
have recalculated $\chi$ after removing small instantons by hand from the
cooled configurations. Here the value at $m_q=0.00625$ has dropped to $19(2)
\times 10^{-5}$, indicating that small instantons which are a lattice artifact
are part of the problem, which suggests that the lattice spacing is too large
to allow complete separation of long and short distance effects on these
lattices. The complete results are shown in table~\ref{tab:top}, and compared
with the theoretical value for the confined region 
\begin{equation}
       \chi = \frac{m_q}{n_f^2}\frac{n_f}{4}\langle\bar{\psi}\psi\rangle
\end{equation}
which is valid for small $m_q$.
We note that, in the hadronic phase at $\beta=5.45$, our measured value is
greater than theory, as expected. In the plasma phase, we have equated theory
to zero since $\langle\bar{\psi}\psi\rangle$ should vanish in the chiral limit
in this chirally symmetric phase. The reason $\chi$ does not, and should not 
vanish in this phase is that the theoretical expression above neglects terms
higher order in $m_q$. In fact, in the plasma phase
\begin{equation}
     \chi \propto m_q^{n_f} = m_q^2
\end{equation}
At $6/g^2=5.55$, which is above the transition for both masses, $\chi=3.0(6)$
for $m_q=0.0125$ and, even with small instantons removed $\chi=2.1(5)$ at
$m_q=0.00625$, a ratio of $\approx 1.4$ rather than the predicted $4$. Further
analysis reveals that at least part of this is due to the finite lattice
spacing which raises the energy of the ``zero mode'' associated with the
instanton to a higher value, comparable in magnitude to the quark mass,
invalidating the derivation of this formula \cite{lagae}.  
\begin{table}[htb]
\begin{tabular}{|l|d|d|d|d|}
         &         &     \multicolumn{3}{c|}{$\chi \times 10^{5}$}        \\
$6/g^2$  &  cools  &     standard     &     broad       &     theory      \\
\hline  
5.450    &   50    &     28.2(3.4)    &     19.1(1.9)   &     11.1(0.5)   \\
5.475    &   25    &     19.8(1.9)    &     12.5(1.2)   &     ---         \\
5.500    &   50    &      5.9(1.1)    &      3.6(0.5)   &     0.0         \\
5.525    &   25    &      5.5(0.9)    &      2.9(0.7)   &     0.0         \\
5.550    &   25    &      4.0(0.9)    &      2.1(0.5)   &     0.0         \\
6.000    &   25    &      0.0         &      0.0        &     0.0         \\
\hline 
\end{tabular}
\caption{Topological susceptibility $\chi$. Standard gives the results obtained
by the cooling method. Broad are the results after removing small instantons,
i.e. those where the maximum value of $|F\tilde{F}|$ on a site is $\geq 2$.}
\label{tab:top}
\end{table}

We end this section with an estimate of the temperature of the transition from
hadronic matter to a quark-gluon plasma. Once we know the critical value of
$6/g^2$, the critical temperature $T_c$ is given by
\begin{equation}
         T_c={1 \over N_t a(g^2_c)}.
\end{equation}
The lattice spacing $a(g^2_c)$ can be determined from measurements of the
hadron spectrum at zero temperature at $6/g^2=6/g^2_c$. Since this has not been
done, we have calculated $a(g^2_c)$, from a value of the $\rho$ mass obtained
by interpolation from simulations at other values of $6/g^2$ and $m_q$,
following Blum et al. \cite{bktg}. This yields $T_c=150(9)$MeV for 
$m_q=0.0125$, $T_c=140(8)$MeV for $m_q=0.00625$, and $T_c=128(9)$ at $m_q=0$.

\section{SCREENING LENGTHS}

In this section we present our measurements of screening masses, the inverse
of screening lengths. We measure the correlations of time-like Wilson lines
which are related to the string tension in hadronic matter, and the Debye
screening length/mass in the quark-gluon plasma, which is the range of influence
of a static colour charge in the plasma. The correlations of spacelike Wilson
lines are also measured. In addition we measure the screening masses for 
excitations with the quantum numbers of low mass hadrons, to determine whether
quark-gluon plasma has hadron-like excitations, or whether it is best described
in terms of quark and gluon excitations.

As described in \cite{part1} , we measure correlations between ``fuzzy''
Wilson/Polyakov loops with zero $x$, $y$, and $t$ momenta, correlated in the z
direction. Again, we find that those which were ``blocked'' the maximum number
of times $(3)$, i.e. maximally ``fuzzy'' were the most correlated. For timelike
Wilson lines our subtracted correlation function should behave at large $z$
separations $Z$ as 
\begin{equation}
      P(Z) = A \{ exp[-\mu Z] + exp[-\mu(N_s-Z)] \}.
\end{equation}
As usual one obtains effective screening masses $\mu(Z)$ (where we have
suppressed the mass dependence) by fitting this form for $P(Z-1)$ and $P(Z)$.
Then $\mu(Z) \rightarrow \mu$, the screening mass, which is closely related 
to the Debye screening mass \cite{ai}. The screening masses so
obtained are tabulated in Table~\ref{tab:debye}.
\begin{table}[htb]
\begin{tabular}{|l|c|c|c|c|}
         & \multicolumn{4}{c|}{Temporal Wilson/Polyakov Line Effective Mass} \\
$6/g^2$  &  $Z$=0--1   & $Z$=1--2  & $Z$=2--3  & $Z$=3--4  \\
\hline
5.450    &  1.10(8)    & 0.84(11)  & 0.96(26)  & 1.35(80)  \\
5.475    &  0.94(4)    & 0.70(6)   & 0.73(15)  & 0.93(46)  \\
5.500    &  0.70(8)    & 0.49(9)   & 0.49(13)  & 0.57(30)  \\
5.525    &  0.77(5)    & 0.60(6)   & 0.66(13)  & 0.86(41)  \\
5.550    &  0.84(7)    & 0.75(15)  & 0.91(40)  &  ---      \\
6.000    &  0.74(4)    & 0.68(6)   & 0.66(16)  & 0.78(41)  \\
\end{tabular}
\caption{Effective masses $\mu(Z)$ from temporal Wilson line correlations.}
\label{tab:debye}
\end{table}
As can be seen, except for the $6/g^2=5.55$ results, our data show some
evidence for a plateau as early as the $Z$=1--2 effective masses ($\mu(2)$), so
we use $\mu(2)$ as our estimate for $\mu$. In the plasma phase, the quantity
of interest $\mu/T$ has the values $3.8(7)$ at $6/g^2=5.5$, $4.8(5)$ at
$6/g^2=5.525$, $6.0(1.2)$ at $6/g^2=5.55$ and $5.4(5)$ at $6/g^2=6.0$, not
significantly different from the $m_q=0.0125$ results.

\begin{figure}[htb]
\epsfxsize=4in
\centerline{\epsffile{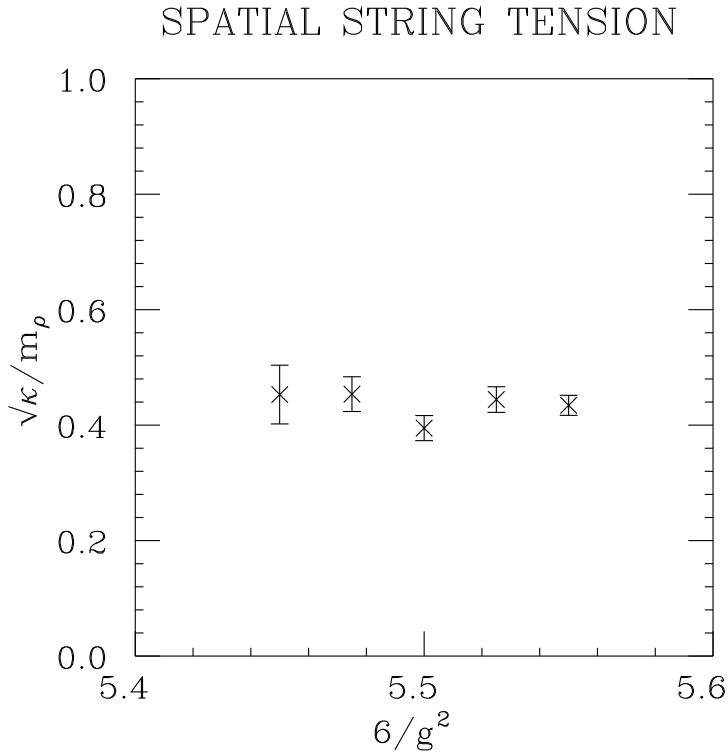}}
\caption{Spatial string tension $\kappa$ as a function of $6/g^2$.
\label{fig:swl}}
\end{figure}

The correlations $C(Z)$ between spatially oriented ``fuzzy'' Wilson/Polyakov
lines, also have an exponential asymptotic behaviour. Unlike the temporal
Wilson/Polyakov lines, the spatial lines do not develop a spatial expectation
value corresponding to a perimeter law in the high temperature phase, but
retain an area law defining a confining theory in both phases. The effective
masses we obtained from C(Z) for maximal blocking are given in
table~\ref{tab:swl}. 
\begin{table}[htb]
\begin{tabular}{|l|c|c|c|c|}
          &  \multicolumn{4}{c|}{Spatial Wilson/Polyakov Line Effective Mass}\\
$6/g^2$   & $Z$=0--1  & $Z$=1--2  & $Z$=2--3  & $Z$=3--4  \\
\hline
5.450     & 2.80(7)   & 1.93(43)  &  ---      &  ---      \\
5.475     & 2.45(6)   & 1.68(22)  & 0.91(83)  &  ---      \\
5.500     & 1.92(5)   & 1.10(12)  &  ---      &  ---      \\
5.525     & 1.74(4)   & 1.19(12)  & 1.28(34)  &  ---      \\
5.550     & 1.53(5)   & 0.97(8)   & 1.09(31)  &  ---      \\
6.000     & 0.51(2)   & 0.41(3)   & 0.39(3)   & 0.40(5)   \\
\end{tabular}
\caption{Effective masses $\mu(Z)$ from spatial Wilson/Polyakov line
         correlation functions}
\label{tab:swl}
\end{table}
We note that these values are very similar to those at $m_q=0.0125$. Taking
$\mu(2)$, the value from fitting $C(Z)$ from $Z=1$ to $Z=2$, as our best
estimate of the mass $\mu$, we can extract a string tension $\kappa=\mu/N_t$.
Fig.~\ref{fig:swl} shows $\sqrt{\kappa}/m_\rho$ as a function of $6/g^2$,
indicating that the string tension appears to be constant in physical units. 
This is in accord with expectations that these spatial Wilson lines should
show confinement at all temperatures \cite{mp}.

Let us now turn to the hadron screening lengths, or rather their inverses, the
hadron screening masses. Not only can measurements of these screening masses
give indications of the nature of excitations in hadronic matter and the 
quark-gluon plasma, but they can also shed light on the nature of chiral
symmetry restoration \cite{dk}. A full understanding of chiral symmetry
realizations at finite temperature requires a study of the flavour singlet
mesons with their disconnected contributions, which is beyond the scope of this
paper. However, as we will see, we can at least address some of these issues. 

For massless quarks, the $SU(2)_A$ component of the $SU(2) \times SU(2)$
chiral symmetry is spontaneously broken, while the $SU(2)_V$ symmetry is
unbroken. The $U(1)_V$ symmetry (baryon/quark number) is conserved, while
the $U(1)_A$ is broken explicitly by the chiral anomaly, as well as 
spontaneously. At the finite temperature transition, the chiral condensate
vanishes, so one expects the $SU(2)_A$ and hence the $SU(2) \times SU(2)$ 
chiral symmetry to be restored. What is less clear, is whether the $U(1)_A$
is restored to give $U(2) \times U(2)$ chiral symmetry \cite{shuryak}

The description of which spatial hadron propagators were considered to get 
our screening masses was given in our first paper, and will not be repeated
here, except to note that wall sources and the $x-y-t$ Coulomb gauge were
employed again, the correlations being measured in the z direction. We also
measured screening masses for hadronic excitations involving valence ``strange''
quarks whose masses are given above. 

\begin{figure}[htb]
\centerline{\hbox{\psfig{figure=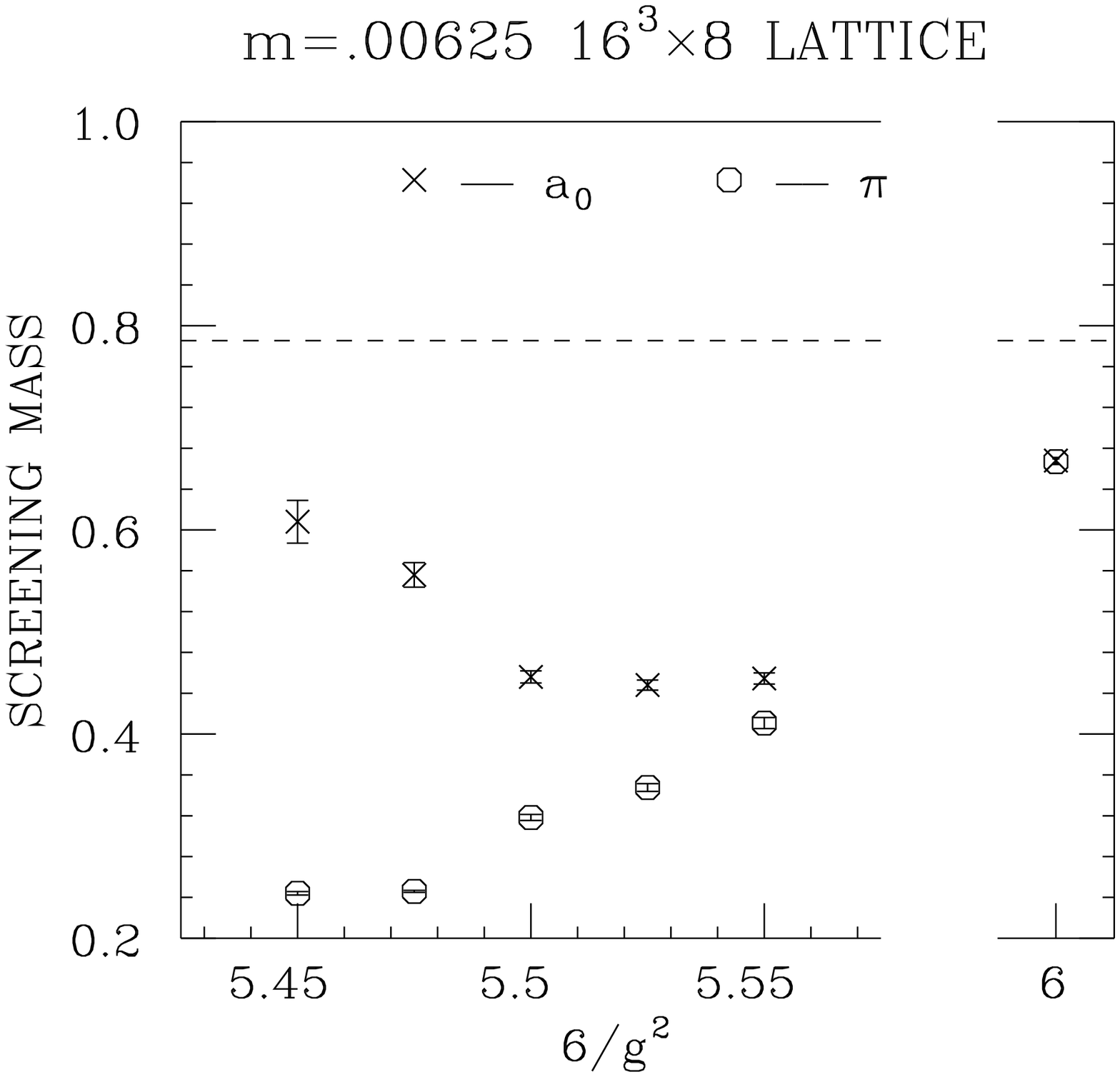,height=8cm,width=8cm},
                  \psfig{figure=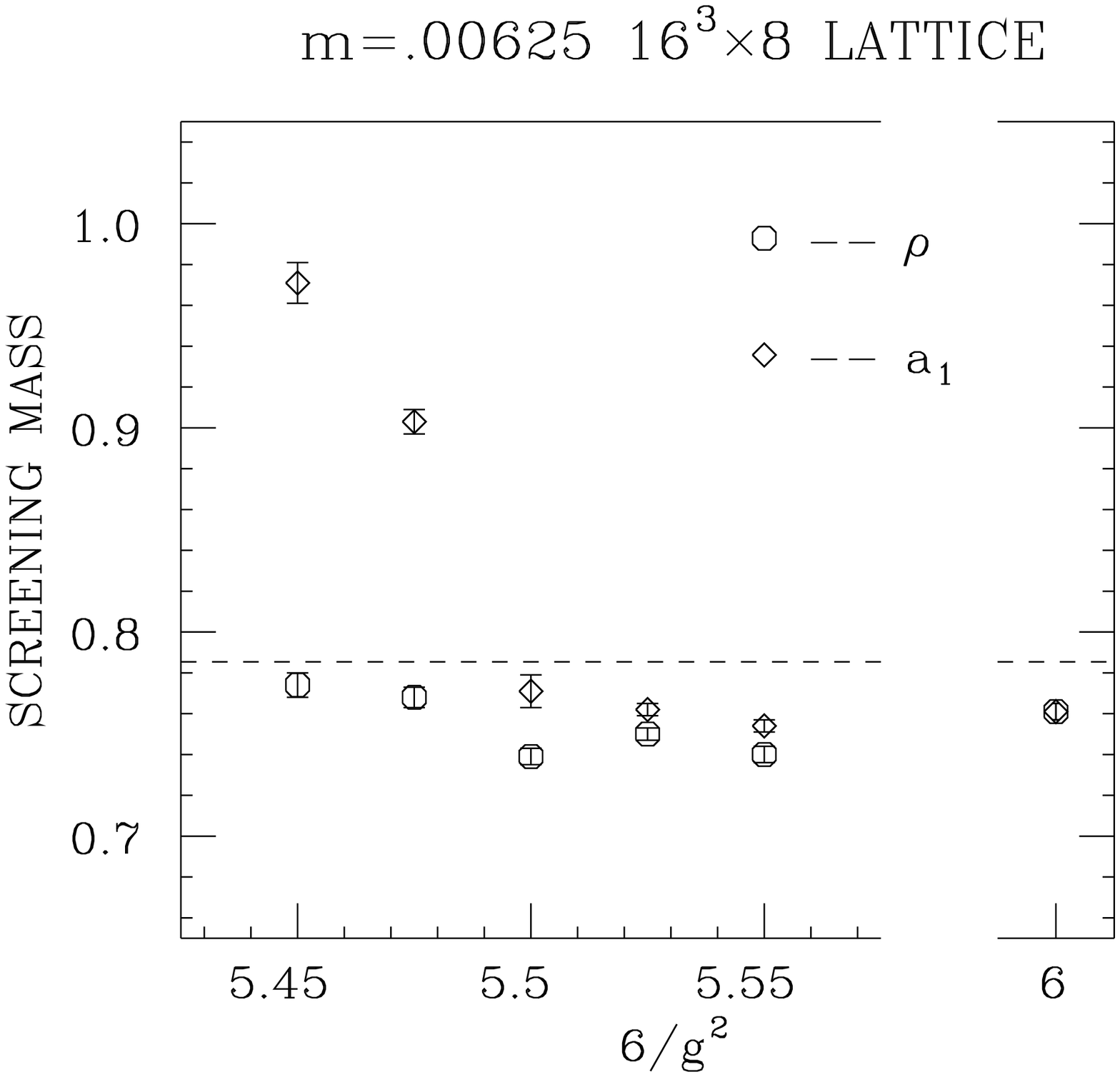,height=8cm,width=8cm}}}
\centerline{(a) \hspace{2.5in} (b)}
\vspace{0.1in}
\centerline{\hbox{\psfig{figure=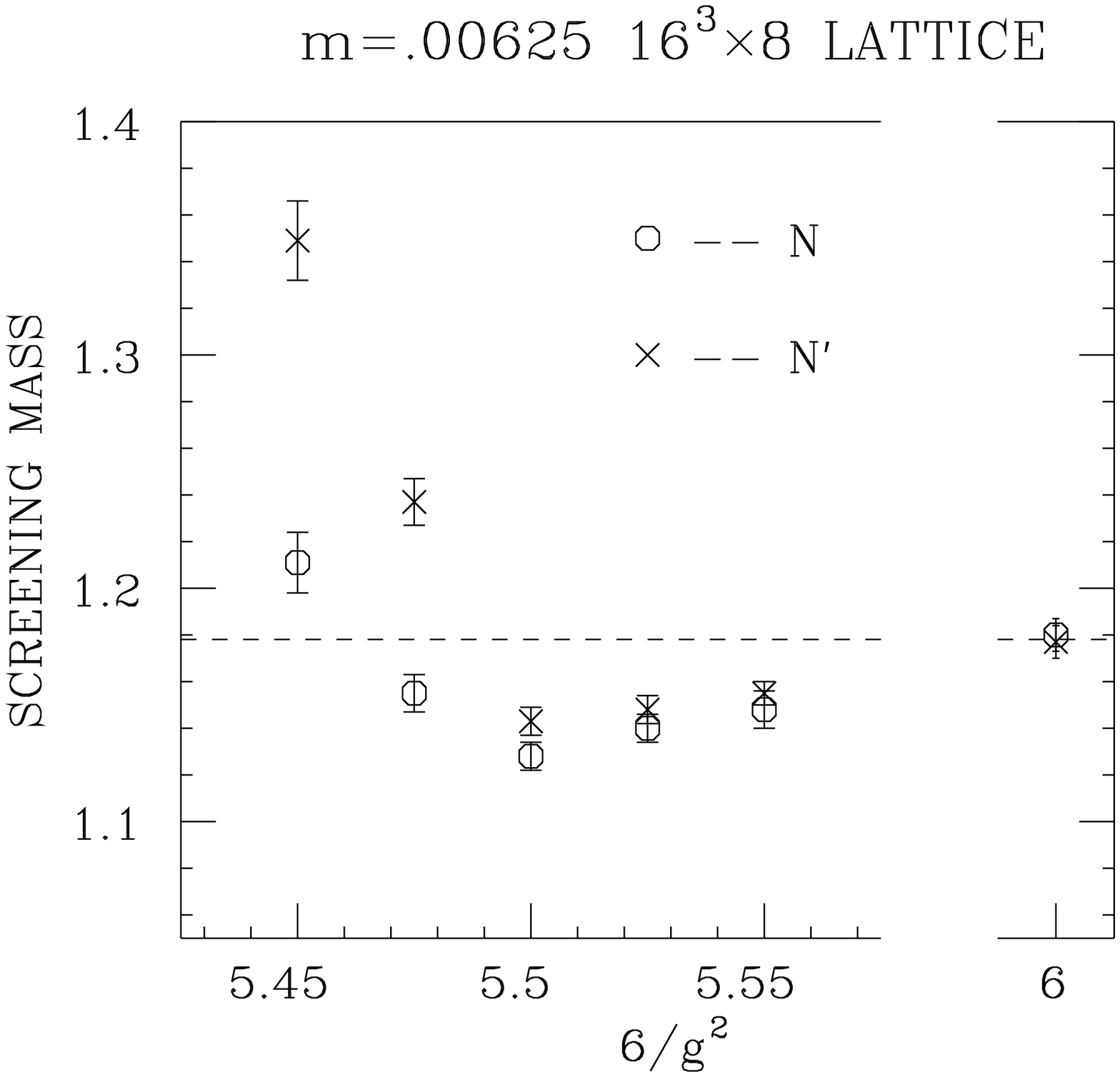,height=8cm,width=8cm},
                  \psfig{figure=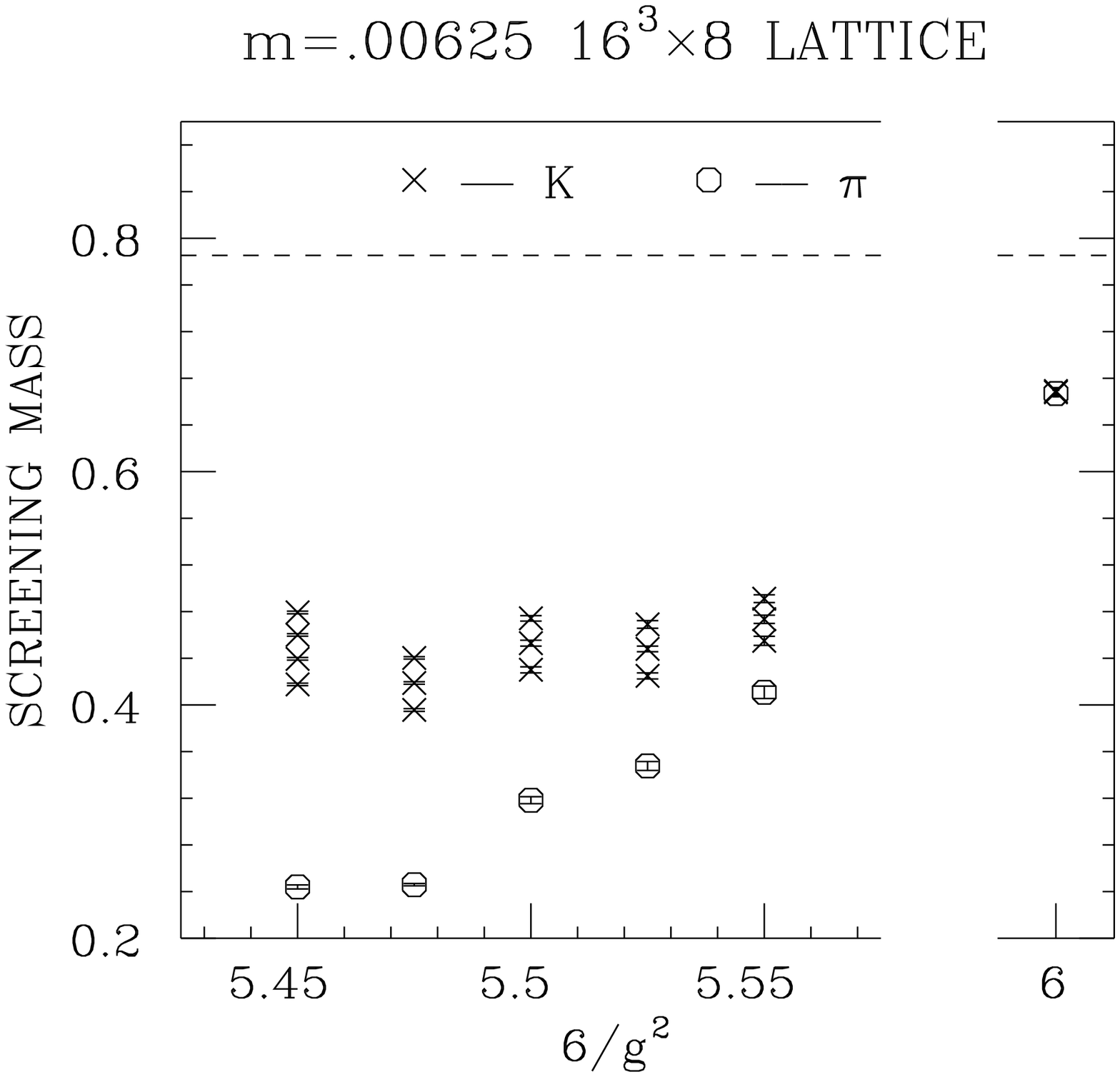,height=8cm,width=8cm}}}
\centerline{(c) \hspace{2.5in} (d)}
\vspace{0.1in}
\caption{Hadronic screening masses as functions of $6/g^2$:
         (a) Scalar states -- $\pi$, $a_0$,
         (b) Vector states -- $\rho$, $a_1$,
         (c) Nucleon states -- $N$, $N'$,
         (d) $K$, $\pi$.
             \label{fig:hadron}}
\end{figure}

Fig.~\ref{fig:hadron}a shows the screening masses of the $\pi$ and the
connected part of the $f_0$ ($\sigma$) propagator -- the $a_0$ propagator.
The error bars reflect only statistical errors. These two masses would become
equal when $U(1)_A$ symmetry is restored. The $\pi$ and $a_0$ appear to have
an appreciable mass splitting just above the transition, only approaching
one another for considerably higher $6/g^2$. The fact that this splitting
appears to be greater than at $m_q=0.0125$ indicates that finite quark mass
is probably not the reason for this splitting. However, since the $a_0$ mass
is difficult to measure we cannot rule out at least part of this splitting
being due to systematic errors. More importantly, since flavour symmetry
breaking is large at these $6/g^2$ values, we cannot rule out the splitting
being due to flavour symmetry breaking. Taken at face value this data suggests
that $U(1)_A$ symmetry is not restored at the transition, but instead is
restored as $T \rightarrow \infty$. Recent measurements of the disconnected
part of the $f_0$ propagator \cite{lagae}, and the susceptibility associated
with it \cite{milc,bklo,christ} indicate that it splits the $f_0$ and $a_0$
masses in the neighbourhood of the transition and gives $f_0$--$\pi$
degeneracy. This supports the scenario where $SU(2)_A$ is restored at the
transition, as expected, but $U(1)_A$ is only restored at somewhat higher
temperatures. 

The $\rho$ and $a_1$ screening masses as a function of $6/g^2$ are presented in
Fig.~\ref{fig:hadron}b. Since these states lie in the same $SU(2) \times SU(2)$
multiplet, their masses should become equal when $SU(2)_A$ is restored. What we
observe is that these screening masses do come together just above the
transition indicating that $SU(2) \times SU(2)$ is restored at the transition,
in the chiral limit. Similarly, Fig.~\ref{fig:hadron}c indicates that the $N$
and $N'$ (negative parity nucleon) become degenerate at (or near) the
transition which is also evidence that chiral $SU(2) \times SU(2)$ symmetry is
restored.

We also note that, just above the transition, the $\rho$ and $a_1$ masses lie 
just below the energy of two free quarks at the lowest Matsubara frequency,
i.e. $2\pi T$, and presumably approach this value from below as 
$T \rightarrow \infty$. Similarly the $N$ and $N'$ masses lie just below the
energy of three free quarks at the lowest Matsubara frequency ($3\pi T$), and
approach this value at high temperatures. Thus these states appear to be
weakly bound above $T_c$. In contrast, both the $\pi$ and the $a_0$ masses are
considerably below $2\pi T$, just above the transition, only approaching this
limit as $T \rightarrow \infty$, indicating that they remain strongly bound,
even in what is usually thought of as the deconfined phase. Thus the 
quark-gluon plasma does appear to support hadronic excitations.

Finally in Fig.~\ref{fig:hadron}d, we present our kaon screening lengths for
the choices of the valence strange quark mass given above. The pion screening
lengths are given for comparison. Note that, the pion screening mass is for
$m_{u,d}=0.00625$ which is considerably larger than the physical quark mass, so
that the pion screening masses will be too large. However, the strange quark
masses are tuned to give reasonable values for the kaon mass, so that kaon
screening lengths we obtained should be reasonable estimates for the physical
kaon screening lengths.

\section{DISCUSSION AND CONCLUSIONS} 

We have extended our previous simulations of the thermodynamics of lattice QCD
with two light staggered quarks down to quark mass $m_q=0.00625$ (lattice
units), in an attempt to probe the chiral limit. At this mass we find a
transition at a critical coupling $6/g_c^2=5.49(2)$ corresponding to a critical
temperature $T_c=140(8)$MeV. Combining this with our previous result at
$m_q=0.0125$, we predict a critical coupling given by $6/g_c^2=5.44(3)$
corresponding to $T_c=128(9)$MeV in the chiral ($m_q=0$) limit. As expected,
there is no evidence for a first order transition. 

The screening length associated with thermal Wilson/Polyakov line correlations
is quantitatively similar to that obtained from our published $m_q=0.0125$
calculations. Thus our conclusion that we should expect $\psi/J$ suppression
for $T \gtrsim 2 T_c$ still stands, while the situation for 
$T_c < T \lesssim 2 T_c$ remains unclear.

We observed that the entropy of strange quarks is $\sim 70\%$ of that of either
$u$ or $d$ quarks, just above the transition. It is to be hoped that 
phenomenology based of this number might shed light on whether there is an
increase in kaon production as the system passes through this transition.

Our study of hadron screening lengths indicates that the $\pi$, $K$ and $a_0$
states remain strongly bound above the transition. Thus there are excitations
with these quantum numbers (and presumably those of the $f_0$), in the
quark-gluon plasma. Such binding is presumably related to the finite string
tension we have observed in our spatial Wilson line correlations. The other
hadronic excitations we observed ($\rho$, $a_1$, $N$ and $N'$) appear to be
only weakly bound above the transition. The relatively large separation of the
$\pi$ and $a_0$ masses above the transition compared with that of the $\rho$
and $a_1$ suggests that these do not become degenerate at the transition, so
that the restored symmetry above the transition is 
$SU(2) \times SU(2) \times U_V(1)$ rather than $U(2) \times U(2)$ as some have
suggested. This is in agreement with recent calculations of the chiral
susceptibility $\chi_m$, and calculations of the $f_0$ mass from the full
(connected + disconnected) propagator. 

Our calculation of the baryon number susceptibility, gives us a glimpse
of physics at finite baryon number density. Calculation of topological
susceptibility at this low mass appears to be strongly affected by the large
lattice spacing at $T_c$, preventing us from reliably extracting its mass
dependence.

For the standard Wilson action which we used, we need a larger value of $N_t$
to remove finite lattice spacing effects. We also need a larger aspect ratio
($N_s/N_t$) to make finite lattice size correction factors closer to one,
reducing the systematic uncertainties in extracting entropy densities, energy
densities and pressures. The larger $N_s$ is also needed to reduce the 
systematic errors involved in extracting screening lengths. Zero temperature
values of the plaquette observable and $\langle\bar{\psi}\psi\rangle$ are
needed to extract partial pressures and energy densities. Measurements closer
together in $6/g^2$ will be necessary if we are to extract the fluctuation
quantities involved in obtaining the equation of state.

\section{ACKNOWLEDGEMENTS}

The computations were carried out on the Connection Machine CM2 at the 
Pittsburgh supercomputer Center, under an NSF grand challenge allocation.
This work was supported by the U.~S. Department of Energy under contract
W-31-109-ENG-38, 
and grants
DE-FG02-91ER-40661, 
DE-FG05-92ER-40742, DE-FG05-85ER250000, 
DE-FG03-95ER-40906, 
and the National Science Foundation under grants
NSF-PHY91--16964, 
NSF-PHY-9503371, 
NSF-PHY92-00148. 
S.~K. is supported by KOSEF through CTP.

\end{document}